\documentclass[aps,reprint]{revtex4-1}
\usepackage{amsmath}
\usepackage{amssymb}
\usepackage{amsthm}
\usepackage{graphicx,subfigure}
\usepackage[bookmarks=false]{hyperref}
\usepackage{bm,bbm}
\hypersetup{colorlinks=true,citecolor=blue,linkcolor=red,%
urlcolor=blue,pdfstartview=FitH,bookmarksopen=true}

\usepackage{algorithm}
\usepackage{algorithmicx}
\usepackage{algpseudocode}

\usepackage[T1]{fontenc}
\usepackage{txfonts}

\DeclareMathOperator{\Tr}{Tr}

\DeclareMathOperator{\diag}{diag}
\newtheorem{theorem}{Theorem}
\newtheorem{proposition}[theorem]{Proposition}

\newtheorem{lemma}[theorem]{Lemma}

\newcommand{\bra}[1]{\langle #1\rvert}
\newcommand{\ket}[1]{\lvert #1\rangle}
\newcommand{\mean}[1]{\langle #1\rangle}

\newcommand{\abs}[1]{\lvert #1\rvert}

\newcommand{\vect}[1]{\bm{#1}}
\renewcommand{\d}{\mathrm{d}}
\newcommand{\I}{\mathbbm{1}}

\newcommand{\maxover}[1][]{\underset{#1}{\text{maximize}}}
\newcommand{\minover}[1][]{\underset{#1}{\text{minimize}}}
\newcommand{\subto}{\text{subject to}}

\begin{document}

\title{Detecting coherence via spectrum estimation}
\author{Xiao-Dong Yu}
\email{Xiao-Dong.Yu@uni-siegen.de}
\affiliation{Naturwissenschaftlich-Technische Fakult\"at, Universit\"at Siegen, 
Walter-Flex-Stra\ss e 3, 57068 Siegen, Germany}
\author{Otfried G\"uhne}
\email{otfried.guehne@uni-siegen.de}
\affiliation{Naturwissenschaftlich-Technische Fakult\"at, Universit\"at Siegen, 
Walter-Flex-Stra\ss e 3, 57068 Siegen, Germany}
\date{\today}
\begin{abstract}
  Coherence is a basic phenomenon in quantum mechanics and considered to be an 
  essential resource in quantum information processing.  Although the quantification 
  of coherence has attracted a lot of interest, the lack of efficient methods to 
  measure the coherence in experiments limits the applications. 
  We address this problem by introducing an experiment-friendly method for 
  coherence and spectrum estimation. This method is based on the theory of 
  majorization and can not only be used to prove the presence of coherence, 
  but also result in a rather precise lower bound of the amount of coherence. 
  As an illustration, we show how to characterize the freezing phenomenon of 
  coherence with only two local measurements for any $N$-qubit quantum systems.  
  Our approach also has other applications in quantum information processing, 
  such as the characterization of distillability and entanglement 
  transformations.
\end{abstract}

\maketitle

\section{Introduction}

Quantum coherence is a fundamental feature of quantum mechanics, describing 
the capability of a quantum state to exhibit quantum interference phenomena. 
Consequently, it is an essential ingredient in quantum information processing 
\cite{Nielsen.Chuang2000}, and plays a central role in emergent fields, such 
as quantum metrology \cite{Giovannetti.etal2004,Toth.Appelaniz2014} 
and quantum thermodynamics \cite{Goold.etal2016}.

The notion of quantum coherence was developed early in quantum optics 
\cite{Glauber1963,Sudarshan1963,Mandel.Wolf1995}, but only in recent years has 
the quantification of coherence been treated rigorously and frameworks for 
quantifying coherence proposed  
\cite{Streltsov.etal2017,Aberg2006,Gour.Spekkens2008,Baumgratz.etal2014,
Marvian.Spekkens2014,Levi.Mintert2014,Girolami2014}.
Notably, in Ref.~\cite{Baumgratz.etal2014} a rigorous framework 
based on the notion of a general resource theory was introduced. 
In this framework, the free states are the incoherent states which 
are diagonal in the incoherent basis, and the free operations are 
incoherent operations whose Kraus operators map the incoherent 
states to incoherent states. Other frameworks were also proposed 
to make the quantification of coherence applicable to various physical 
situations \cite{Aberg2006,Marvian.Spekkens2014,Yadin.etal2016,
Peng.etal2016,deVicente.Streltsov2017,Chitambar.Gour2016,Yu.etal2016b}.
The main difference between these frameworks is that they have different 
notions of free operations.

Based on these frameworks, several coherence 
measures have been proposed, such as the relative entropy of coherence 
\cite{Aberg2006,Baumgratz.etal2014}, the $l_1$ norm of coherence 
\cite{Baumgratz.etal2014}, the geometric measure of coherence 
\cite{Streltsov.etal2015}, the robustness of coherence 
\cite{Napoli.etal2016,Piani.etal2016}, and others 
\cite{Yuan.etal2015,Winter.Yang2016,Du.etal2015,Zhu.etal2017,Rana.etal2017,Bu.etal2017}.  
These coherence measures make it possible to {\it quantitatively} study the 
role of coherence in various physical contexts. Further important properties of 
coherence, such as the distillation of coherence 
\cite{Winter.Yang2016,Chitambar.etal2016}, the relation between coherence and 
quantum correlations \cite{Streltsov.etal2015,Yao.etal2015,Xi.etal2015, 
Radhakrishnan.etal2016,Chitambar.Hsieh2016,Streltsov.etal2016,Ma.etal2016}, and 
the freezing phenomenon of coherence \cite{Bromley.etal2015,Yu.etal2016},
can be studied based on the coherence measures that have been proposed.

While many theoretical works have been devoted to the quantification of 
coherence, only a few results have been examined in experiments 
\cite{Silva.etal2016,Wang.etal2017,Yu.etal2017}.  One important reason for this 
situation is the fact that few methods are known to obtain the coherence 
measures in experiments 
\cite{Ringbauer.etal2018,Smith.etal2017,Zhang.etal2018,Biswas.etal2017}, and 
these methods either use the normal witness technique, or require complicated 
experiment settings or numerical optimizations. The lack of efficient and 
scalable methods for coherence detection severely limits the applications of 
coherence measures.

To overcome this situation and to make the quantification of coherence 
a common tool for quantum information processing, it is of paramount 
importance to improve the evaluation of coherence in experiments. In this work, 
we address this problem by developing an efficient method to witness and 
estimate the coherence of quantum systems based on spectrum estimation of the
density matrix. With this method, we can not only witness the presence of coherence,
but also obtain a good estimation of coherence of quantum systems with only few 
measurements. As an illustration, we show how to use our method to characterize 
the freezing phenomenon of coherence with only two local measurements for any 
$N$-qubit quantum systems. Our approach relies on the mathematical theory of 
the majorization lattice and can, as we explain, also be used for other 
problems in quantum information processing. As the majorization theory is also 
widely-used in physics, statistics, and economics 
\cite{Horodecki.Oppenheim2013,Bengtsson.Zyczkowski2017,Marshall.etal2011}, our 
approach may have potential applications beyond quantum information.


\section{Resource theory of coherence}

In the resource theory of coherence, the free states are incoherent states 
$\mathcal{I}$, defined as $\delta=\sum_ip_i\ket{i}\bra{i}$, where $\{\ket{i}\}$ 
represents a fixed reference basis, known as the incoherent basis. The 
definition of free operations within the resource theory of coherence is not 
unique. Several approaches have been proposed based on different physical or 
mathematical considerations \cite{Streltsov.etal2017}. With these definitions 
of free states and free operations, the frameworks for quantifying coherence 
are constructed from the general resource theory 
\cite{Brandao.Gour2015,Coecke.etal2016} and many coherence measures are 
proposed \cite{Streltsov.etal2017}.

In the text, we focus on the estimation of relative entropy of coherence, 
defined as $C_r(\rho)=\min_{\delta\in\mathcal{I}}S(\rho\|\delta)$, where 
$S(\rho\|\delta)=\Tr(\rho\log_2\rho-\rho\log_2\delta)$ is the relative entropy.
This is not only because the relative entropy of coherence is a legitimate 
coherence measure in all of the proposed frameworks.  More importantly, the 
estimation of relative entropy of coherence can be used for estimating many 
other quantities in quantum information processing, such as the distillable 
coherence, the intrinsic randomness, and the secrete key rate in quantum key 
distribution \cite{Winter.Yang2016,Yuan.etal2015,Cao.etal2016,
Devetak.Winter2005,Ma.etal2018}. The applications of our method to other 
coherence measures are discussed in the appendixes. Mathematically, the 
relative entropy of coherence also admits the closed form
\begin{equation}
  C_r(\rho)=S(\rho_d)-S(\rho),
  \label{eq:realtiveEntropyCoherence}
\end{equation}
where $S$ is the von Neumann entropy and $\rho_d$ is the diagonal part of 
$\rho$ in the incoherent basis.

\section{Majorization and the majorization lattice}

A probability distribution $\vect{a}=(a_1,a_2,\dots,a_n)$ is said to majorize 
a probability distribution $\vect{b}=(b_1,b_2,\dots,b_n)$ (written as 
$\vect{a}\succ\vect{b}$), if it satisfies 
$\sum_{i=1}^ka_i^\downarrow\ge\sum_{i=1}^kb_i^\downarrow$ for all 
$k=1,2,\dots,n$, where $\vect{a}^\downarrow$ ($\vect{b}^\downarrow$) is the 
probability distribution with the same components as $\vect{a}$ ($\vect{b}$), 
but sorted in descending order. Hereafter, we assume that the components of 
a probability distribution are already sorted in descending order and all 
vectors are column vectors, unless stated otherwise.

Majorization imposes an important constraint between measurement results and 
the spectrum of quantum states, shown in the following well-known fact 
\cite{Bengtsson.Zyczkowski2017}.

\begin{lemma}
  Let $\rho$ be a quantum state in an $n$-dimensional Hilbert space with 
  spectrum $\vect{\lambda}=(\lambda_1,\lambda_2,\dots,\lambda_n)$.
  \begin{enumerate}
    \item[a.]
      If we perform a von Neumann measurement on the quantum state and get the 
      probability distribution of measurement outcomes 
      $\vect{p}=(p_1,p_2,\dots,p_n)$, then $\vect{p}\prec\vect{\lambda}$.
    \item[b.]
      The above condition is also sufficient in the sense that if a probability 
      distribution $\vect{p}=(p_1,p_2,\dots,p_n)$ fulfills that $\vect{p}\prec 
      \vect{\lambda}$, then there exists a von Neumann measurement 
      $\{P_i\}_{i=1}^n$ such that $p_i=\Tr(P_i\rho)$.
  \end{enumerate}
  \label{lem:majorization}
\end{lemma}

For the estimation of coherence, we also take advantage of the strict 
Schur-concavity of the Shannon entropy $S$. Mathematically, it says that if 
$\vect{a}\prec\vect{b}$ and $\vect{a}\ne\vect{b}$, then 
$S(\vect{a})>S(\vect{b})$.

Compared with majorization, the theory of the majorization lattice is much 
less-known. Given two distributions $\vect{a}$ and $\vect{b}$ which are not 
comparable in the sense of majorization, one may ask whether there is 
a ``smallest'' distribution majorizing both of them. In fact, such 
a distribution does exist and is called the majorization join. Similarly, the 
``largest'' distribution majorized by both $\vect{a}$ and $\vect{b}$ is called 
the majorization meet. Rigorously, a probability distribution $\vect{c}$ is 
called the majorization join (meet) of $\vect{a}$ and $\vect{b}$ if it 
satisfies:
(i) $\vect{c}\succ\vect{a},\vect{b}$ ($\vect{c}\prec\vect{a},\vect{b}$), and
(ii) $\vect{c}\prec\vect{\tilde{c}}$ ($\vect{c}\succ\vect{\tilde{c}}$) for any 
$\vect{\tilde{c}}$ that satisfies $\vect{a},\vect{b}\prec\vect{\tilde{c}}$ 
($\vect{a},\vect{b}\succ\vect{\tilde{c}}$) \cite{Cicalese.Vaccaro2002}.
The majorization join and meet of $\vect{a}$ and $\vect{b}$ are usually denoted 
$\vect{a}\vee\vect{b}$ and $\vect{a}\wedge\vect{b}$, respectively. The basic 
process for constructing majorization join (meet) is quite simple.  The 
necessary and sufficient condition for $\vect{\tilde{c}}\prec\vect{a},\vect{b}$ 
is $\sum_{i=1}^k\tilde{c}_i\le\min\{\sum_{i=1}^ka_i,\sum_{i=1}^kb_i\}$.  Let 
$\sum_{i=1}^kc_i=\min\{\sum_{i=1}^ka_i,\sum_{i=1}^kb_i\}$; then 
$\vect{c}\prec\vect{a},\vect{b}$ and $\vect{c}\succ\vect{\tilde{c}}$, i.e., 
$\vect{c}=\vect{a}\wedge\vect{b}$, if $c_k$ are in descending order. For 
majorization meet, this construction is sufficient. For majorization join, the 
vector $\vect{c}$ constructed from 
$\sum_{i=1}^kc_i=\max\{\sum_{i=1}^ka_i,\sum_{i=1}^kb_i\}$ may not be in 
descending order. Some further flattening operations may be needed. See 
Ref.~\cite{Cicalese.Vaccaro2002} or Appendix \ref{sec:majorizationLattice} for 
more details.

\section{Single-partite systems}

From Eq.~\eqref{eq:realtiveEntropyCoherence}, we can easily see that the 
coherence of the quantum system can be revealed by the diagonal part of the 
quantum state $\rho$, i.e., the vector  $\vect{d}=(d_1,d_2,\dots,d_n)$, and the 
spectrum of $\rho$, i.e., the vector 
$\vect{\lambda}=(\lambda_1,\lambda_2,\dots,\lambda_n)$.  In experiments, one 
can easily determine the diagonal part of the quantum state $\rho$ by just 
measuring in the incoherent basis. It remains to determine the spectrum of 
$\rho$, although the eigenbasis is unknown.

Suppose that $\rho$ is incoherent; then we have $\vect{d}=\vect{\lambda}$.  
According to Lemma~\ref{lem:majorization}, the probability distribution 
$\vect{p}=(p_1,p_2,\dots,p_n)$ from any von Neumann measurement must satisfy 
$\vect{p}\prec\vect{\lambda}=\vect{d}$. So, if we perform a measurement and get 
the probability distribution $\vect{p}$ such that $\vect{p}\nprec\vect{d}$, 
then we can immediately assert that $\rho$ is coherent.  Conversely, if $\rho$ 
is coherent, there always exists a measurement $\{P_i\}_{i=1}^n$ such that the 
probability distribution of the outcomes $\vect{p}$ satisfies 
$\vect{p}\nprec\vect{d}$. For example, we can choose $\{P_i\}_{i=1}^n$ to be 
the eigenbasis of $\rho$.

The previous discussion implies that we can always prove the presence of 
coherence by showing that the probability distribution of the outcomes of some 
measurement is not majorized by the probability distribution of the measurement 
outcomes in the incoherent basis. This provides a qualitative method for 
witnessing coherence. However, contrary to the case of
entanglement and separability, the set of incoherent states is of measure $0$ 
in the state space, so a mere statement about the presence of coherence is of 
limited value. Instead, a quantitative method, giving an estimate of the amount 
of coherence, is desirable. So the problem arises what can one say about the 
amount of coherence in a quantum system, if the probability distributions 
$\vect{p}$  and $\vect{d}$ with  $\vect{p}\nprec \vect{d}$
are known.

To answer this question, we use the strict Schur-concavity of the Shannon 
entropy, which implies that $S(\rho)=S(\vect{\lambda})\le S(\vect{p})$.  
Consequently, we have a lower bound of coherence,
\begin{equation}
  C_r(\rho)\ge \max\{0, S(\vect{d})-S(\vect{p})\}.
  \label{eq:roughLowerBound}
\end{equation}
Still, this bound may not be strictly positive, even if we can 
conclude from $\vect{p}\nprec \vect{d}$ that the quantum system contains 
coherence. For example, consider the qutrit state 
$\ket{\varphi}=(\ket{0}+\ket{1})/\sqrt{2}$ and the measurement in the basis
$\{(\ket{0}+\ket{1}+\ket{2})/{\sqrt{3}},
(\ket{0}+\omega\ket{1}+\omega^2\ket{2})/{\sqrt{3}},
(\ket{0}+\omega^2\ket{1}+\omega\ket{2})/{\sqrt{3}}\}$
with $\omega=e^{\frac{2\pi i}{3}}$.
This gives $\vect{d}=({1}/{2}, {1}/{2},0)$,
$\vect{p}=(\frac{2}{3},\frac{1}{6},\frac{1}{6})$,
and $\vect{p}\nprec\vect{d}$, but the lower bound from 
Eq.~\eqref{eq:roughLowerBound} is still $0$, as $S(\vect{d})-S(\vect{p})<0$.

The main idea of solving the problem is that we can construct the ``smallest'' 
probability distribution that majorizes both $\vect{p}$ and $\vect{d}$, i.e., 
the majorization join $\vect{p}\vee\vect{d}$. According to 
Lemma~\ref{lem:majorization}, we have that $\vect{d}\prec\vect{\lambda}$ and 
$\vect{p}\prec\vect{\lambda}$. From the definition of majorization join, we get 
that $\vect{d}\vee\vect{p}\prec\vect{\lambda}$ and 
$\vect{d}\prec\vect{d}\vee\vect{p}$. Furthermore, $\vect{p}\nprec\vect{d}$ 
implies that $\vect{d}\ne\vect{d}\vee\vect{p}$. Hence, the strict 
Schur-concavity of the Shannon entropy implies that 
$C_r(\rho)=S(\vect{d})-S(\vect{\lambda})\ge 
S(\vect{d})-S(\vect{d}\vee\vect{p})>0$. These results are summarized in the 
following proposition.

\begin{proposition}
  If the probability distribution of the outcomes of a von Neumann measurement 
  $\vect{p}$ is not majorized by diagonal entries of the quantum state 
  $\vect{d}$, i.e., $\vect{p}\nprec\vect{d}$, then the quantum system contains
  coherence.  Furthermore, a nonzero lower bound of the relative entropy 
  of coherence is given by
  \begin{equation}
    C_r(\rho)\ge S(\vect{d})-S(\vect{d}\vee\vect{p}).
    \label{eq:majorizationEstimation}
  \end{equation}
  \label{thm:majorizationEstimation}
\end{proposition}

The lower bound provided by Proposition~\ref{thm:majorizationEstimation} is
calibration-free in the sense that we do not resort to the actual form of the 
measurement basis for measuring the probability distribution $\vect{p}$.  The 
benefit of this feature is that the bound in 
Eq.~\eqref{eq:majorizationEstimation} is robust to the errors in implementing 
the measurement. In addition, the lower bound is also tight in this sense.
To show this, we only need to prove that there is a quantum state $\rho$ such 
that (i) $\vect{d}$ is its diagonal part; (ii) $\vect{p}$ is the probability 
distribution of outcomes of some von Neumann measurement $P$; and (iii) 
$\vect{d}\vee\vect{p}$ is its spectrum.  Suppose 
$\vect{d}\vee\vect{p}=(c_1,c_2,\dots,c_n)$ and let 
$\tilde{\rho}=\sum_{i=1}^nc_i\ket{i}\bra{i}$. According to Lemma 
\ref{lem:majorization}, there are two bases $\{\ket{\varphi_i}\}_{i=1}^n$ and 
$\{\ket{\psi_i}\}_{i=1}^n$ such that 
$\bra{\varphi_i}\tilde{\rho}\ket{\varphi_i}=d_i$ and 
$\bra{\psi_i}\tilde{\rho}\ket{\psi_i}=p_i$, as $\vect{d}\prec\vect{d} 
\vee\vect{p}$ and $\vect{p}\prec\vect{d}\vee\vect{p}$. Then we can choose the 
state $\rho=U\tilde{\rho}U^\dagger$ and the von Neumann measurement 
$P=\{U\ket{\psi_i}\bra{\psi_i}U^\dagger\}_{i=1}^n$, where 
$U=\sum_{i=1}^n\ket{i}\bra{\varphi_i}$. We can easily verify that $\rho$ and 
$P$ fulfill the three conditions above, and hence the lower bound in 
Proposition \ref{thm:majorizationEstimation} is tight.

The method represented in Eq.~\eqref{eq:majorizationEstimation} can also be 
generalized to the case where many measurements are performed. Suppose that 
$\vect{p}_1,\vect{p}_2,\dots,\vect{p}_k$ are the probability distributions of 
measurement outcomes of $k$ different measurements, then we can take advantage 
of the majorization join of all $\vect{p}_1,\vect{p}_2,\dots,\vect{p}_k$ and 
$\vect{d}$ to estimate the lower bound of coherence
\begin{equation}
  C_r(\rho)\ge S(\vect{d})-S(\vect{d}\vee\vect{p}_1\vee\dots\vee\vect{p}_k).
\end{equation}
As $\vect{d}\vee\vect{p}_1\vee\dots\vee\vect{p}_k\succ
\vect{d}\vee\vect{p}_1\vee\dots\vee\vect{p}_{k-1}\succ\dots\succ
\vect{d}\vee\vect{p}_1$, the Schur-concavity of the Shannon entropy implies 
that we can successively improve the estimation of coherence. This method is 
quite efficient in practice by employing adaptive strategies. More details are 
given in Appendix \ref{sec:adaptive}. In actual experiments, one usually has 
some expectations or predictions concerning the state of the quantum system.  
This may also be used to choose the measurement and provide better bounds on 
coherence.

\section{Multi-partite systems}

The incoherent basis of multi-partite systems is usually defined based on the 
tensor product of the incoherent bases for each subsystem 
\cite{Streltsov.etal2017}. The difference between the multi-partite case and 
the single-partite case is that since usually only local measurements are 
allowed for multi-partite quantum systems in experiments, we cannot get the 
probability distribution $\vect{p}$ for an entangled basis efficiently. As 
a compromise, we resort to the estimation method of $\vect{p}$. In entanglement 
detection theory, many efficient methods have been developed to estimate the 
fidelity $\bra{\varphi}\rho\ket{\varphi}$ with local measurements 
\cite{Guehne.Toth2009}. Applying these methods to the entangled basis 
$\{\ket{\varphi_i}\}_{i=1}^n$, we can estimate the probability distribution 
$\vect{p}=(p_1,p_2,\dots,p_n)$, where the components  
$p_i=\bra{\varphi_i}\rho\ket{\varphi_i}$ may not be in descending order.  
Usually, the estimations can be expressed as linear constraints 
$A\vect{p}\ge\vect{\alpha}$ and $B\vect{p}=\vect{\beta}$, where $A$ and $B$ are 
matrices, $\vect{\alpha}$ and $\vect{\beta}$ are vectors, and ``$\ge$'' denotes 
the component-wise comparison. For example, in the theory of witnesses for 
graph states this is the case \cite{Toth.Guehne2005b}.

Let $X$ denote the feasible set, i.e., $X=\{\vect{p}\mid 
A\vect{p}\ge\vect{\alpha}, B\vect{p}=\vect{\beta}\}$. Using the transitivity of 
the majorization relation, it directly follows that 
$\vect{\lambda}\succ\vect{d}\vee(\bigwedge_{\vect{p}\in X}\vect{p})$, where 
$\bigwedge_{\vect{p}\in X}\vect{p}$ is the majorization meet of all probability 
distributions in $X$. Then the Schur-concavity of Shannon entropy implies that
\begin{equation}
  C_r(\rho)\ge S(\vect{d})-S(\vect{d}\vee({\textstyle\bigwedge}_{\vect{p}\in 
  X}\vect{p})).
  \label{eq:majorizationEstimationMulti}
\end{equation}
The main difficulty in calculating the bound in 
Eq.~\eqref{eq:majorizationEstimationMulti} is the majorization meet of the 
infinite number of probability distributions in $X$. In the following, we show 
that this problem can be converted into a linear program, for which efficient 
algorithms exist \cite{Boyd.Vandenberghe2004}.

Suppose that $\hat{\vect{p}}\equiv\bigwedge_{\vect{p}\in X}\vect{p}= 
(\hat{p}_1,\hat{p}_2,\dots,\hat{p}_n)$, then $\hat{p}_k=s_k-s_{k-1}$, where 
$s_0=0$ and $s_k$ can be written as the convex optimization problem,
$\min_{\vect{p}\in X}\sum_{i=1}^kp_i^{\downarrow}$,
for $k=1,2,\dots,n$. Notably, here the components of $\vect{p}$ may not be in 
descending order, so the ``$\downarrow$'' is necessary for $p_i^{\downarrow}$.  
By combining the minimax theorem with duality techniques, we convert the 
optimization $s_k=\min_{\vect{p}\in X}\sum_{i=1}^kp_i^{\downarrow}$
to the linear program
\begin{equation}
  \begin{aligned}
    &\maxover[\vect{\mu},\vect{\nu}]\quad
	  && \vect{\alpha}^T\vect{\mu}+\vect{\beta}^T\vect{\nu}\\
    &\subto
	  && \vect{0}\le A^T\vect{\mu}+B^T\vect{\nu}\le\vect{1}\\
    &     && \vect{1}^TA^T\vect{\mu}+\vect{1}^TB^T\vect{\nu}=k\\
    &     && \vect{\mu}\ge\vect{0},
  \end{aligned}
  \label{eq:meetDual}
\end{equation}
where $\vect{\mu}$ and $\vect{\nu}$ are vectors that have the same dimension as 
$\vect{\alpha}$ and $\vect{\beta}$, respectively, and $\vect{0}$ and $\vect{1}$ 
are vectors with all components being $0$ and $1$, respectively. See Appendix 
\ref{sec:MajLP} for the proof.

The estimation method based on Eq.~\eqref{eq:majorizationEstimationMulti} is 
quite efficient in practice. For example, we have tested the optimization 
problem for systems of up to $10$ qubits with $513$ equality constraints and 
$1024$ inequality constraints using the \textsc{cvxpy} package 
\cite{Diamond.Boyd2016}.  The vector $\hat{\vect{p}}=\bigwedge_{\vect{p}\in 
X}\vect{p}$ with $1024$ components can be determined within $30$ min in 
a common laptop.

Before discussing examples, we note that our method also has other applications 
in quantum physics. First, it may be applied to the majorization criterion for 
distillability of quantum states \cite{Nielsen.Kempe2001, Hiroshima2003}. This 
criterion states that for an undistillable state the eigenvalues of the global 
state $\rho_{AB}$ are majorized by the eigenvalues of the reduced state 
$\rho_A$. At first sight, it seems that state tomography is required for 
checking this relation, but our methods provide a way to circumvent this. As 
already mentioned, from some local measurements one can typically obtain linear 
constraints on the eigenvalues of the global state and reduced state. Using 
Eq.~(\ref{eq:meetDual}), we can compute the majorization meet of the possible 
global eigenvalues and the majorization join of the local ones. If they violate 
the relation mentioned above, the state must be distillable. This concept may 
be generalized to other separability criteria based on majorization 
\cite{Augusiak.Stasinska2009}.

Second, majorization of the Schmidt coefficients of pure states provides 
a necessary and sufficient condition for the state transformations under local 
operations and classical communication in the resource theory of entanglement 
\cite{Nielsen1999}. Thus, our methods can be used to obtain the common resource 
states that can generate a whole subclass of entangled states. This concept may 
also be generalized to the case of approximate state transformations 
\cite{Vidal1999,Vidal.etal2000,Bosyk.etal2017}.

Third, as our method provides an estimation of the spectrum, it can be used to 
estimate any Schur-convex or Schur-concave quantities, such as the purity, 
Tsallis entropy, and R\'enyi entropy. The method can also be used for 
estimating other coherence measures \cite{Baumgratz.etal2014,Napoli.etal2016,
Zhao.etal2018,Regula.etal2018,Fang.etal2018,Zhao.etal2018b}.
More details are reported in Appendix \ref{sec:otherEst}, in which some novel 
relations between different coherence measures are also proposed.

\section{Characterizing the freezing of coherence}

The freezing of coherence means that the coherence of the quantum system 
(quantified by some coherence measure) is not affected by noise.  When the 
freezing of coherence is independent of the choice of measures, it is called 
universal freezing \cite{Bromley.etal2015,Yu.etal2016}. Especially, in 
Ref.~\cite{Yu.etal2016}, it is shown that under a strictly incoherent channel, 
the universal freezing of coherence occurs if and only if the relative entropy 
of coherence is frozen.  This implies that if we can witness the freezing of 
relative entropy of coherence, we can assure that the coherence of the quantum 
system is completely unaffected by noise.

One of the most important examples of universal freezing of coherence is the
$N$-qubit GHZ state $\rho_0=(\ket{0}^{\otimes N}+\ket{1}^{\otimes 
N})/\sqrt{2}$, in the local bit-flip channel $\Lambda^{\otimes N}$, where 
$\Lambda(\rho)=(1/2+e^{-\gamma t}/2)\rho+(1/2-e^{-\gamma 
t}/2)\sigma_x\rho\sigma_x$ and $\gamma$ is a parameter that represents the 
strength of the noise. Direct calculations show that the state at time $t$ is 
of the form
$\rho_t=\sum_{l}p_l^\pm\ket{\varphi_l^\pm}\bra{\varphi_l^\pm}$,
where $\ket{\varphi_l^\pm}$ is the GHZ basis, i.e.,
$\ket{\varphi_l^\pm}=\ket{\varphi_{l_1l_2\ldots l_N}^\pm}=(\ket{l_1l_2\ldots 
l_N}\pm\ket{\bar{l}_1\bar{l}_2\ldots\bar{l}_N})/\sqrt{2}$,
with $l_1=0$, $l_{i\neq1}=0,1,$ and $\bar{l}_i=1-l_i$. For convenience, we 
split the time-dependent probabilities $\vect{p}=(p_l^\pm)$ into two parts, 
$\vect{p}=\vect{p}^+\oplus\vect{p}^-=(p_l^+)\oplus(p_l^-)$.

In order to get an accurate estimation of coherence, we would like to perform 
a GHZ-basis measurement on the quantum system. Since the GHZ-basis measurement 
is highly entangled, it is not easy to get the exact probability distribution 
in experiments. As a compromise, we choose to estimate the probability 
distribution with the following two local measurement settings,
\begin{equation}
  \mathcal{X}={\textstyle\bigotimes}_{i=1}^N\sigma^{(i)}_x,\quad
  \mathcal{Z}={\textstyle\bigotimes}_{i=1}^N\sigma^{(i)}_z.
  \label{eq:GHZstablizer}
\end{equation}
An advantage of this measurement setting is that we can get not only the 
expectation values $\mean{\mathcal{X}}$ and $\mean{\mathcal{Z}}$, but also all 
$\mean{\mathcal{Z}_S}$ with $\mathcal{Z}_S=\bigotimes_{i\in S}\sigma^{(i)}_z$, 
for any subset $S$ of $\{1,2,\dots,N\}$. Furthermore, all $\mathcal{X}$ and 
$\mathcal{Z}_E$ are diagonal in the GHZ basis, when $E$ is a nonempty subset of 
$\{1,2,\dots,N\}$ with an even number of elements. Hence, from  
$\mean{\mathcal{X}}$ and $\mean{\mathcal{Z}_E}$, together with the condition 
that $\vect{p}$ is a probability distribution, we can get $2^{N-1}+1$ linear 
equalities and $2^N$ linear inequalities for the estimation of the probability 
distribution $\vect{p}$. As all the constraints are linear equalities or 
inequalities, Eqs.~\eqref{eq:majorizationEstimationMulti} and 
\eqref{eq:meetDual} can be applied immediately to estimate coherence. See 
Appendix \ref{sec:example} for more details about the construction of the 
linear program and Appendix \ref{sec:symmetrization} for how to simplify the 
linear program using the symmetrization technique \cite{Toth.etal2010}.

In the ideal case where the fidelity of the initial GHZ state is $1$, we have 
that $\mean{\mathcal{X}}=1$ at any time. This implies that 
$\vect{p}^-=\vect{0}$ at any time.  Then the $2^{N-1}-1$ independent equalities 
from $\mean{\mathcal{X}_E}$ completely determine the probability distribution 
$\vect{p}^+$. Hence, we get $\bigwedge_{\vect{p}\in 
X}\vect{p}=\vect{p}^+\oplus\vect{0}$. Additionally, the measurement 
$\mathcal{Z}$ will also give us the diagonal part of the quantum state 
$\vect{d}=\frac{1}{2}\vect{p}^+\oplus\frac{1}{2}\vect{p}^+$. Then 
Eq.~\eqref{eq:majorizationEstimationMulti} implies that $C_r(\rho_t)\ge 1=
C_r(\rho_0)$. Thus, we prove the freezing of coherence with only two local 
measurements given in Eq.~\eqref{eq:GHZstablizer}.

\begin{figure}
  \centering
  \includegraphics[width=.98\linewidth]{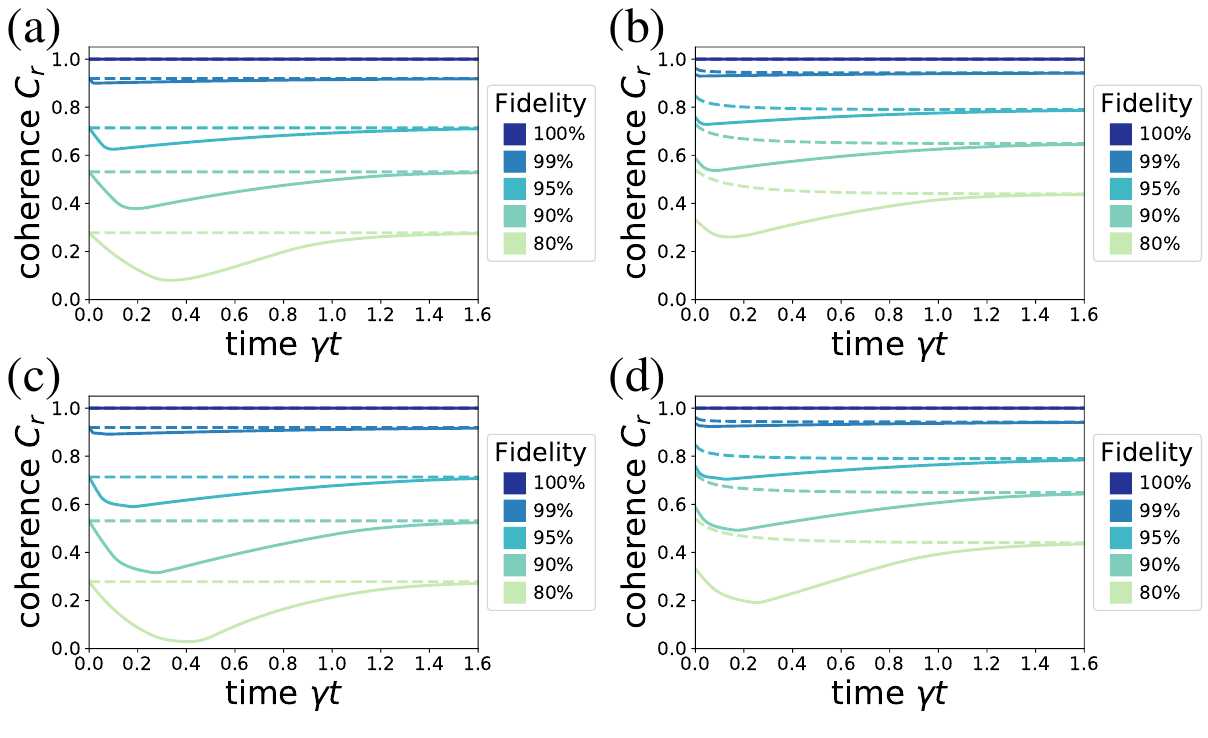}
  \caption{
    Characterizing the freezing of coherence. We consider the situation where 
    the preparation of the initial state is affected by dephasing and 
    depolarizing noise in the three-qubit [(a) and (b), respectively] and 
    four-qubit [(c) and (d), respectively] cases. Dashed lines are the actual 
    values of coherence and solid lines are the estimations of coherence with 
    our method.  Lines of different colors represent initial states with 
    different fidelities.
  }
  \label{fig:freezing}
\end{figure}

In actual experiments, the fidelity of the initial GHZ state is always strictly 
smaller than $1$. We consider two common noise models for experiments, 
dephasing noise and depolarizing noise, where the detailed noise models are 
shown in Appendix \ref{sec:freezing}. Note that in the case of depolarizing 
noise, the freezing of coherence no longer occurs, but the coherence is still 
very resistant to noise. In both cases, our method can still prove the 
resistance of coherence to noise, as illustrated in Fig.~\ref{fig:freezing}.

\section{Conclusions}

In this work, we propose an efficient method for coherence detection based on 
the majorization lattice. This method is efficient in two senses: first, the 
number of measurements needed is quite small, which is friendly to experiments; 
second, the optimization process is only linear programming, which can be 
efficiently solved in practice. As an illustration, we show that we can 
characterize the freezing phenomenon of coherence with only two local 
measurements.  We hope that this work can not only promote the verification of 
various results on the quantification of coherence, but also promote the 
application of coherence measures to quantum information experiments. Moreover, 
as the majorization theory is widely-used in many fields besides quantum 
information, such as physics, statistics, and economics, our efficient method 
for characterizing the majorization lattice also has many other potential 
applications.

\appendix

\section{Majorization lattice}
\label{sec:majorizationLattice}

For simplicity, we restrict our discussion to probability distributions.  All 
the discussions can be naturally generalized to the general case. We first 
recall the definition of majorization.  A probability distribution 
$\vect{a}=(a_1,a_2,\dots,a_n)$ is said to majorize probability distribution 
$\vect{b}=(b_1,b_2,\dots,b_n)$ written as $\vect{a}\succ\vect{b}$, if they 
satisfy $\sum_{i=1}^ka_i^\downarrow\ge\sum_{i=1}^kb_i^\downarrow$ for all 
$k=1,2,\dots,n$, where $\vect{a}^\downarrow$ ($\vect{b}^\downarrow$) is the 
vector with the same components as $\vect{a}$ ($\vect{b}$), but sorted in 
descending order. As in the text, we assume that the components of 
a probability distribution are already sorted in descending order, unless 
otherwise stated.

The majorization lattice deals with the ``smallest'' (``largest'') probability 
distribution that majorizes (is majorized by) two probability distributions, 
$\vect{a}$ and $\vect{b}$. More formally, a probability distribution $\vect{c}$ 
is called the majorization join of $\vect{a}$ and $\vect{b}$, if it satisfies 
that
\begin{enumerate}
  \item
    $\vect{c}\succ\vect{a},\vect{b}$;
  \item
    $\vect{c}\prec\vect{\tilde{c}}$ for any $\vect{\tilde{c}}$ that satisfies 
    $\vect{a},\vect{b}\prec\vect{\tilde{c}}$.
\end{enumerate}
Similarly, a probability distribution $\vect{c}$ is called the majorization 
meet of $\vect{a}$ and $\vect{b}$, if it satisfies that
\begin{enumerate}
  \item
    $\vect{c}\prec\vect{a},\vect{b}$;
  \item
    $\vect{c}\succ\vect{\tilde{c}}$ for any $\vect{\tilde{c}}$ that satisfies 
    $\vect{a},\vect{b}\succ\vect{\tilde{c}}$.
\end{enumerate}
From the definition, we can easily check that the majorization join and meet 
are unique. We denote the majorization join and meet of $\vect{a}$ and 
$\vect{b}$ as $\vect{a}\vee\vect{b}$ and $\vect{a}\wedge\vect{b}$, 
respectively.  From the definition, we can also easily prove that the 
majorization join and meet satisfy the commutativity and associativity, e.g.,
\begin{enumerate}
  \item
    $\vect{p}_1\vee\vect{p}_2=\vect{p}_2\vee\vect{p}_1$;
  \item
    $(\vect{p}_1\vee\vect{p}_2)\vee\vect{p}_3 
    =\vect{p}_1\vee(\vect{p}_2\vee\vect{p}_3)$.
\end{enumerate}
Hence, we can simply denote the majorization join of $m$ probability 
distributions $\vect{p}_i,~i=1,2,\dots,m$ as 
$\vect{p}_1\vee\vect{p}_2\vee\dots\vee\vect{p}_m$ or 
$\bigvee_{i=1}^m\vect{p}_i$ and the majorization join of all probability 
distributions in a set $X$ (finite or infinite) as $\bigvee_{\vect{p}\in 
X}\vect{p}$. Similar notations can also be used for majorization meet.

The basic process for the construction of the majorization meet 
$\bigwedge_{\vect{p}\in X}\vect{p}$ is quite simple.  The necessary and 
sufficient condition for $\vect{\tilde{c}}\prec\vect{p}$ for all $\vect{p}\in X$ is 
$\sum_{i=1}^k\tilde{c}_i\le\inf_{\vect{p}\in X}\sum_{i=1}^kp_i$.  Hence if 
$\sum_{i=1}^kc_i=\inf_{\vect{p}\in X}\sum_{i=1}^kp_i$ and $c_k$ are in 
descending order, then $\vect{c}\prec\vect{p}$ and 
$\vect{c}\succ\vect{\tilde{c}}$, i.e., $\vect{c}=\bigwedge_{\vect{p}\in 
X}\vect{p}$. In the case of majorization meet, this is always possible. Just 
let
\begin{equation}
  c_k=s_k-s_{k-1},
  \label{eq:majorizationLattice}
\end{equation}
for $k=1,2,\dots,n$, where
\begin{equation}
  s_0=0 \text{ and } s_k=\inf_{\vect{p}\in X}\sum_{i=1}^kp_i.
  \label{eq:majorizationMeetSk}
\end{equation}
Then for any $\vect{p}\in X$, we have 
$\sum_{i=1}^{k+1}p_i+\sum_{i=1}^{k-1}p_i\le 2\sum_{i=1}^kp_i$, as $p_k$ are in 
descending order. This implies that $\inf_{\vect{p}\in 
X}\sum_{i=1}^{k+1}p_i+\inf_{\vect{p}\in X}\sum_{i=1}^{k-1}p_i\le 
2\inf_{\vect{p}\in X}\sum_{i=1}^kp_i$, i.e., $s_{k+1}+s_{k-1}\le 2s_{k}$. Thus 
$c_{k+1}=s_{k+1}-s_k\le s_k-s_{k-1}=c_k$, i.e., components of $\vect{c}$ are in 
descending order.

For the majorization join $\bigvee_{\vect{p}\in X}\vect{p}$, the construction is 
a little bit more complicated. The probability distribution obtained from 
Eq.~\eqref{eq:majorizationLattice} with
\begin{equation}
  s_0=0 \text{ and } s_k=\sup_{\vect{p}\in X}\sum_{i=1}^kp_i,
  \label{eq:majorizationJoinSk}
\end{equation}
may not be in descending order. Some further flattening operations may be 
needed as shown in the following algorithm (steps 2-7). The main idea is that, 
for all $k=1,2,\dots,n$, the flattening operation never decreases $c_k$ and 
always preserves the relation
\begin{equation}
  \sum_{i=1}^kc_i\le\sum_{i=1}^k\tilde{c}_i,
  \label{eq:flattening}
\end{equation}
where $\vect{\tilde{c}}$ is any probability distribution such that 
Eq.~\eqref{eq:flattening} holds initially, i.e., $\sup_{\vect{p}\in 
X}\sum_{i=1}^kp_i\le\sum_{i=1}^k\tilde{c}_i$. This property of the flattening 
operation can be easily checked with some basic calculations 
\cite{Cicalese.Vaccaro2002}.

\begin{algorithm}[H]
  \caption{Majorization join}
  \begin{algorithmic}[1]
    \State
    Let $s_0=0$ and $s_k=\sup_{\vect{p}\in X}\sum_{i=1}^kp_i$, for 
    $k=1,2,\dots,n$;
    \State
    Let $c_k=s_k-s_{k-1}$, for $k=1,2,\dots,n$;
    \For {$k=3,\dots,n$,}
    \If{$c_k>c_{k-1}$,}
	\State Find the largest $l<k$ such that
	$\frac{1}{k-l+1}\sum_{i=l}^kc_i\le c_{l-1}$;
	\State Update each of $c_l,c_{l+1},\dots,c_k$ to 
	$\frac{1}{r-l+1}\sum_{k=l}^rc_k$;
      \EndIf
    \EndFor
  \end{algorithmic}
\end{algorithm}

\noindent\textbf{Examples:}
\begin{enumerate}
  \item
    Consider the probability distributions 
    $\vect{p}=(\frac{2}{3},\frac{1}{6},\frac{1}{6})$ and 
    $\vect{d}=(\frac{1}{2},\frac{1}{2},0)$, which are studied in the main text.  
    Then $\vect{s}=(\frac{2}{3},1,1)$ in step 1 and 
    $\vect{c}=(\frac{2}{3},\frac{1}{3},0)$ in step 2. In this case, we do not 
    need to do the flattening operation, because $c_3\le c_2$.  Hence, we get
    $\vect{p}\vee\vect{d}=(\frac{2}{3},\frac{1}{3},0)$.
  \item
    Let $\vect{a}=(\frac{2}{3},\frac{1}{9},\frac{1}{9},\frac{1}{9})$ and 
    $\vect{b}=(\frac{1}{2},\frac{1}{4},\frac{1}{4},0)$; then 
    $\vect{s}=(\frac{2}{3},\frac{3}{4},1,1)$. In step 2, we get that
    $\vect{c}=(\frac{2}{3},\frac{1}{12},\frac{1}{4},0)$. In this case, we have 
    $c_3>c_2$.  Hence, we need to find the largest $l<3$ such that
    $\frac{1}{4-l}\sum_{k=l}^3c_k\le c_{l-1}$, which gives $l=2$. Then update 
    both $c_2$ and $c_3$ to $\frac{1}{2}(c_2+c_3)=\frac{1}{6}$, which gives 
    $\vect{c}=(\frac{2}{3},\frac{1}{6},\frac{1}{6},0)$. No further operation is 
    needed, because $c_4\le c_3$. Hence, the final result is 
    $\vect{a}\vee\vect{b}=(\frac{2}{3},\frac{1}{6},\frac{1}{6},0)$.
\end{enumerate}

%
%

\section{Adaptive measurements for estimating \\
spectrum and coherence}
\label{sec:adaptive}

If there is no prediction or expectation of the quantum state, instead of 
choosing the measurements randomly, one can use the following strategy. The 
basic idea of this strategy is quite simple. The previous measurements 
constrain the state, and we choose the next measurement basis such that there 
exists a possible state that is diagonal in this basis. Rigorously, let the 
previous measurements be $\{U_1,U_2,\dots,U_k\}$, where $U$ denotes the 
measurement in the basis $\{U\ket{i}\}_{i=1}^n$ and $U_1=\I$ is the measurement 
in the incoherent basis; then the $(k+1)$-th measurement is chosen such that 
there exists a quantum state $\rho$ satisfying that
\begin{equation}
  \begin{aligned}
    &U_{k+1}\rho U_{k+1}^\dagger \text{ is diagonal, and}\\
    &\Delta(U_i^\dagger\rho U_i)=\vect{p}_i, \text{ for } i=1,2,\dots,k,
  \end{aligned}
  \label{eq:compatibileMeasurement}
\end{equation}
where $\Delta(X)$ is the diagonal part of $X$ and $\vect{p}_i$ are the 
probability distributions obtained from the previous measurements. This is 
equivalent to the feasibility problem
\begin{equation}
  \begin{aligned}
    &\text{find}                  && \rho \\
    &\subto                       && \Delta(U_i\rho U_i)=\diag(\vect{p}_i)\\
    &                             && \rho\ge 0.
  \end{aligned}
  \label{eq:feasibilitySDP}
\end{equation}
Suppose that $\hat{\rho}$ satisfies Eq.~\eqref{eq:feasibilitySDP}; then we can 
choose the $U_{k+1}$ such that $U_{k+1}\hat{\rho}U_{k+1}^\dagger$ is diagonal.  
The choice of $\hat{\rho}$ is not unique, unless the previous measurements are 
informationally complete. Here, we try to maximize the majorization of the 
spectrum of $\rho$. As the majorization is only a partial order, we can at most 
expect some local maximum. The algorithm here is based on the simple 
observation that $\vect{a}\succ\vect{b}$ if and only if 
$\sum_{i=1}^nc_ia_i\ge\sum_{i=1}^nc_ib_i$ for all $\vect{c}$, where the 
components of $\vect{a}$, $\vect{b}$, and $\vect{c}$ are all in descending 
order. By choosing $\vect{c}=(n,n-1,\dots,1)$, the observation leads to 
a see saw algorithm; i.e., one first randomly chooses an initial $\hat{H}$, 
then alternatively solves the following two optimization problems until 
convergence:
\begin{equation}
  \begin{aligned}
    &\maxover[\rho]               && \Tr(\hat{H}\rho) \\
    &\subto                       && \Delta(U_i\rho U_i)=\diag(\vect{p}_i)\\
    &                             && \rho\ge 0,
  \end{aligned}
  \label{eq:seeSawRho}
\end{equation}
\begin{equation}
  \begin{aligned}
    &\maxover[H,U]                && \Tr(H\hat{\rho}) \\
    &\subto                       && H=U^\dagger DU\\
    &                             && U\in \mathrm{SU}(n),
  \end{aligned}
  \label{eq:seeSawH}
\end{equation}
where $\hat{H}$ is the optimal solution of Eq.~\eqref{eq:seeSawH}, $\hat{\rho}$ 
is the optimal solution of Eq.~\eqref{eq:seeSawRho}, and 
$D=\diag(n,n-1,\dots,1)$.  Equation~\eqref{eq:seeSawRho} is a semidefinite 
program, which can be efficiently solved \cite{Boyd.Vandenberghe2004}. The 
solution of Eq.~\eqref{eq:seeSawH} is given by $\hat{H}=\hat{U}^\dagger 
D\hat{U}$, where $\hat{U}$ satisfies that 
$\hat{U}\hat{\rho}\hat{U}^\dagger=\diag(\vect{\lambda}(\hat{\rho}))$ and 
$\vect{\lambda}(\hat{\rho})$ is the spectrum of $\hat{\rho}$ whose components 
are in descending order.

\begin{figure}
  \centering
  \includegraphics[width=.98\linewidth]{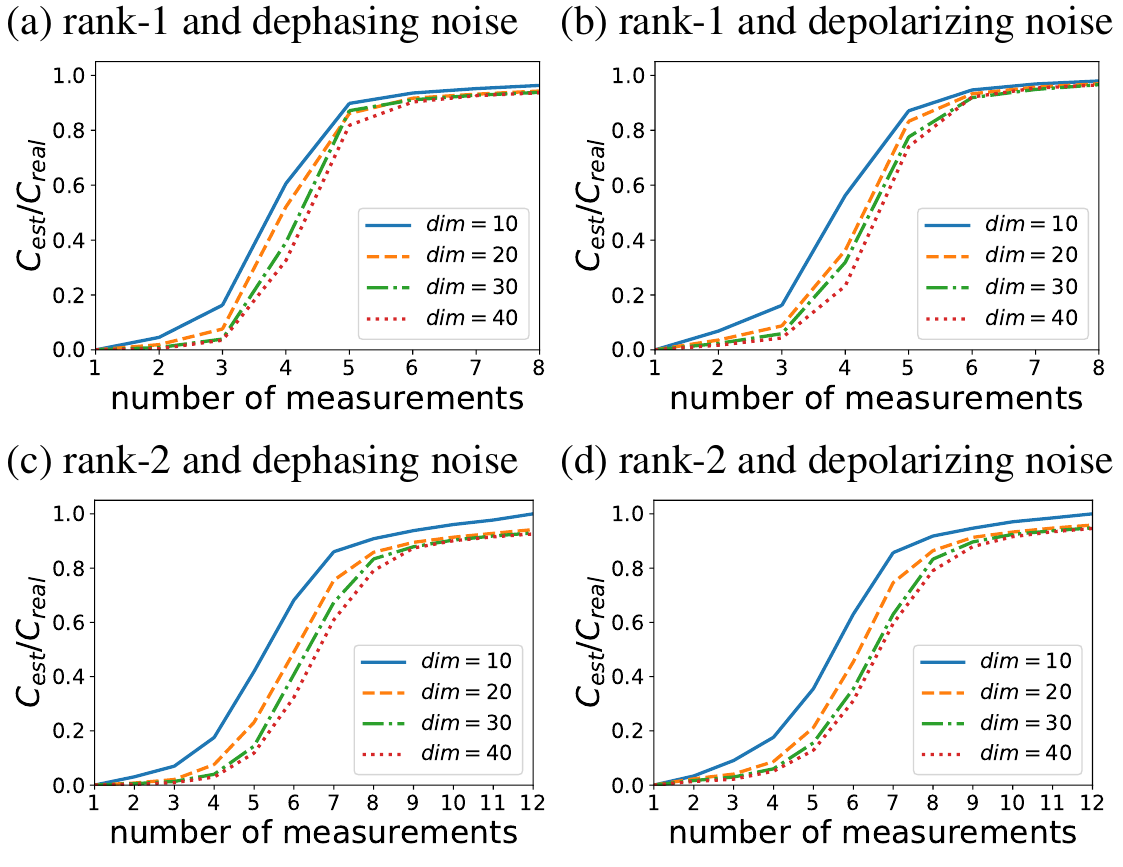}
  \caption{Estimation of coherence with adaptive measurements. The quantum 
    system is in some low-rank state and affected by noise. For the simulation, 
    we choose two common types of noise in experiments, dephasing noise 
    $(1-\varepsilon)\rho+\varepsilon\Delta(\rho)$ and depolarizing (white) 
    noise $(1-\varepsilon)\rho+\frac{\varepsilon}{d}\I$, with noise strength 
    $\varepsilon=0.2$ and randomly generated state $\rho$. Curves show the 
    average ratio between the estimation of coherence ($C_{est}$) and the real 
    value of coherence ($C_{real}$) of $100$ randomly generated rank-$1$ and 
    rank-$2$  states $\rho$ that affected by the dephasing and depolarizing 
  noise.}
  \label{fig:adaptive}
\end{figure}

In the worst case, our method requires $n+1$ measurements to get the coherence 
of an $n$-dimensional quantum state, which is the same performance as the 
quantum state tomography. However, the method can perform much better in 
practice.  For example, if the quantum state is some low-rank quantum state and 
affected by noise, our method provides a significant improvement compared to 
the tomography methods as illustrated in Fig.~\ref{fig:adaptive}.  The 
numerical results show that with a few measurements, we can get a rather 
precise estimation of the coherence, and precision of the estimation barely 
decreases as the dimension goes from $10$ to $40$. We note that our method 
requires no prior information of the quantum state or the noise.

\section{Majorization join and meet over linear constraints}
\label{sec:MajLP}

In this Appendix, we consider a special case of majorization join and meet of 
an infinite number of probability distributions, the majorization join and meet 
over linear constraints.

We first consider the case of majorization meet, i.e., $\bigwedge_{\vect{p}\in 
X}\vect{p}$, where $X=\{\vect{p}\mid A\vect{p}\ge\vect{\alpha}, 
B\vect{p}=\vect{\beta}\}$.  Here $A$ and $B$ are matrices, $\vect{\alpha}$ and 
$\vect{\beta}$ are vectors, and ``$\ge$'' denotes the component-wise 
comparison.  Note that the components of $\vect{p}$ may not be in descending 
order in this case. The conditions that $\vect{p}$ is a probability 
distribution, i.e., $\vect{p}\ge\vect{0}$ and $\vect{1}^T\vect{p}=1$, where 
$\vect{0}$ and $\vect{1}$ are vectors with all components being $0$ and $1$, 
respectively, are already included in the constraints.  According to Appendix 
\ref{sec:majorizationLattice}, $\bigwedge_{\vect{p}\in X}\vect{p}\equiv 
(\hat{p}_1,\hat{p}_2,\dots,\hat{p}_n)$ is given by $\hat{p}_k=s_k-s_{k-1}$, 
where $s_0=0$ and $s_k$ can be written as the convex optimization problem
\begin{equation}
  \begin{aligned}
    &\minover[\vect{p}]\quad && \sum_{i=1}^kp_i^{\downarrow}\\
    &\subto                  && A\vect{p}\ge\vect{\alpha}\\
    &                        && B\vect{p}=\vect{\beta},
  \end{aligned}
  \label{eq:meetPrimalA}
\end{equation}
for $k=1,2,\dots,n$. As the components of $\vect{p}$ in $X$ may not be in 
descending order, the ``$\downarrow$'' is necessary for $p_k^{\downarrow}$ in 
Eq.~\eqref{eq:meetPrimalA}. Now, we prove that the convex optimization is 
equivalent to the linear program
\begin{equation}
  \begin{aligned}
    &\maxover[\vect{\mu},\vect{\nu}]\quad && \vect{\alpha}^T\vect{\mu}
			     +\vect{\beta}^T\vect{\nu}\\
    &\subto
	  && \vect{0}\le A^T\vect{\mu}+B^T\vect{\nu}\le\vect{1}\\
    &     && \vect{1}^TA^T\vect{\mu}+\vect{1}^TB^T\vect{\nu}=k\\
    &     && \vect{\mu}\ge\vect{0},
  \end{aligned}
  \label{eq:meetDualA}
\end{equation}
where $\vect{\lambda}$ and $\vect{\nu}$ are vectors whose dimensions dependent 
on the numbers of inequality and equality constraints in 
Eq.~\eqref{eq:meetPrimalA}.

Let $Y_k=\{\vect{y}\mid\vect{0}\le\vect{y}\le\vect{1},~\vect{1}^T\vect{y}=k\}$; 
then $\sum_{i=1}^kp_i^{\downarrow}=\max_{\vect{y}\in Y_k}\vect{y}^T\vect{p}$.  
Thus we get $\min_{\vect{p}\in X}\sum_{i=1}^kp_i^{\downarrow}=\min_{\vect{p}\in 
X}\max_{\vect{y}\in Y_k}\vect{y}^T\vect{p}$. As both $X$ and $Y_k$ are compact 
and convex, and $\vect{y}^T\vect{p}$ is a continuous convex-concave (actually 
bilinear) function on $(\vect{p},\vect{y})$, the von Neumann minimax theorem 
implies that we can exchange optimizations $\min$ and $\max$, i.e.,
\begin{equation}
  \min_{\vect{p}\in X}\sum_{i=1}^kp_i^{\downarrow}=\min_{\vect{p}\in 
  X}\max_{\vect{y}\in Y_k}\vect{y}^T\vect{p}=\max_{\vect{y}\in 
  Y_k}\min_{\vect{p}\in X}\vect{y}^T\vect{p}.
  \label{eq:biLP}
\end{equation}
Now we consider the first optimization $\min_{\vect{p}\in 
X}\vect{y}^T\vect{p}$, i.e.,
\begin{equation}
  \begin{aligned}
    &\minover[\vect{p}]\quad && \vect{y}^T\vect{p}\\
    &\subto                  && A\vect{p}\ge\vect{\alpha}\\
    &                        && B\vect{p}=\vect{\beta},
  \end{aligned}
  \label{eq:primalLP}
\end{equation}
which is a linear program, and hence the strong duality always holds. Thus 
$\min_{\vect{p}\in X}\vect{y}^T\vect{p}$ equals to
\begin{equation}
  \begin{aligned}
    &\maxover[\vect{\mu},\vect{\nu}]\quad
	  && \vect{\alpha}^T\vect{\mu}+\vect{\beta}^T\vect{\nu}\\
    &\subto
	  && A^T\vect{\mu}+B^T\vect{\nu}=\vect{y}\\
    &     && \vect{\mu}\ge\vect{0}.
  \end{aligned}
  \label{eq:dualLP}
\end{equation}
Combining Eqs.~\eqref{eq:biLP} and \eqref{eq:dualLP}, we get the final linear 
program in Eq.~\eqref{eq:meetDualA}.

Finally, we briefly discuss how to calculate the majorization join over linear 
constraints i.e., $\bigvee_{\vect{p}\in X}\vect{p}$, where $X=\{\vect{p}\mid 
A\vect{p}\ge\vect{\alpha}, B\vect{p}=\vect{\beta}\}$. According to Appendix 
\ref{sec:majorizationLattice}, the main process is to maximize 
$\sum_{i=1}^kp^\downarrow_i$ over $X$.  As $\sum_{i=1}^kp^\downarrow_i$ is 
a convex function on $\vect{p}$, the maximization is always achieved on the 
extreme points. As all the constraints are linear, finding the extreme points 
is equivalent to finding all the vertices of the polytope expressed by $X$, for 
which efficient algorithms exist \cite{Avis.Fukuda1992}. Thus, we transform the 
majorization join over linear constraints to the majorization join of finite 
probability distributions, which can be directly solved by the method shown in 
Appendix \ref{sec:majorizationLattice}.

\section{Estimation of other coherence measures}
\label{sec:otherEst}

In this Appendix, we show that our method can also be used to estimate other 
coherence measures, such as the $l_2$ norm of coherence, $l_1$ norm of 
coherence, robustness of coherence, and two coherence measures used for 
one-shot coherence manipulations.  Notably, this Appendix also contains some 
novel relations between different coherence measures.

First, we consider the $l_2$ norm of coherence, which is a coherence measure 
under genuine incoherent operations
\cite{Baumgratz.etal2014,deVicente.Streltsov2017}.
The $l_2$ norm of coherence is defined as
\begin{equation}
  C_{l_2}(\rho)=\sum_{i\ne 
  j}\abs{\rho_{ij}}^2=\Tr(\rho^2)-\Tr(\rho_d^2)=S_L(\vect{d})-S_L(\vect{\lambda}),
  \label{eq:l2NormCoherence}
\end{equation}
where $S_L(\vect{p})=1-\sum_{i=1}^np_i^2$ is called the Tsallis-2 entropy or 
linear entropy. As $S_L$ is also Schur-concave, we can get that
\begin{equation}
  C_{l_2}(\rho)\ge S_L(\vect{d})-S_L(\vect{d}\vee\vect{p}).
  \label{eq:estimationCl2}
\end{equation}

Second, we show how to use the majorization technique to estimate the 
$l_1$ norm of coherence \cite{Baumgratz.etal2014}
\begin{equation}
  C_{l_1}=\sum_{i\ne j}\abs{\rho_{ij}},
  \label{eq:l1NormCoherence}
\end{equation}
from $C_{l_2}$ and $\vect{d}$. We first study the case where the exact value of 
$C_{l_2}$ is known, and it is generalized to the case where only an estimated 
value is known later. To this end, we consider the optimization problem
\begin{equation}
  \begin{aligned}
    &\minover[\rho_{ij}]\quad     && 2\sum_{i<j}\abs{\rho_{ij}}\\
    &\subto                  && 2\sum_{i<j}\abs{\rho_{ij}}^2=C_{l_2}\\
    &                        && \abs{\rho_{ij}}^2\le d_id_j,
  \end{aligned}
  \label{eq:Cl1fromCl2}
\end{equation}
where $\abs{\rho_{ij}}^2\le d_id_j$ follows from the fact that $\rho$ is 
positive semidefinite. Now, we view $(2\abs{\rho_{ij}}^2/C_{l_2})_{i<j}$ as 
a probability distribution $\vect{v}=(v_k)_{k=1}^{n(n-1)/2}$.  Correspondingly, 
denote $(2d_id_j/C_{l_2})_{i<j}$ as a vector $\vect{u}=(u_k)_{k=1}^{n(n-1)/2}$.  
Without loss of generality, we assume that $u_k$ are in descending order. Then 
the optimization in Eq.~\eqref{eq:Cl1fromCl2} can be written as 
\begin{equation}
  \begin{aligned}
    &\minover[v_k]\quad     && \sqrt{2C_{l_2}}\sum_{k=1}^{n(n-1)/2}\sqrt{v_k}\\
    &\subto                 && \sum_{k=1}^{n(n-1)/2}v_k=1\\
    &                       && 0 \le v_k\le u_k.
  \end{aligned}
  \label{eq:Cl1fromCl2simplified}
\end{equation}
The objective function in Eq.~\eqref{eq:Cl1fromCl2simplified} is 
a Schur-concave function. Hence we can use $\hat{\vect{v}}=\bigvee_{v\in 
R}\vect{v}$ to estimate the lower bound, where $R$ is the feasible region in 
Eq.~\eqref{eq:Cl1fromCl2simplified}. In this special case, 
$\hat{\vect{v}}=\bigvee_{v\in R}\vect{v}$ can be evaluated analytically. By 
performing the algorithm in Appendix \ref{sec:majorizationLattice}, we can 
easily see that
\begin{equation}
  \begin{aligned}
    s_k&=\sum_{l=1}^ku_l \quad && \text{ for } k\le M,\\
    s_k&=1 \quad && \text{ for } k>M,\\
  \end{aligned}
  \label{eq:Cl1JoinSum}
\end{equation}
where $M$ is the largest integer such that $\sum_{l=1}^Mu_l\le 1$. Then we get 
that
\begin{equation}
  \begin{aligned}
    \hat{v}_k&=u_k \quad && \text{ for } k\le M,\\
    \hat{v}_k&=1-\sum_{l=1}^Mu_l \quad && \text{ for } k=M+1,\\
    \hat{v}_k&=0 \quad && \text{ for } k>M+1.\\
  \end{aligned}
  \label{eq:Cl1Join}
\end{equation}
As $u_k$ are in descending order, $\hat{v}_k$ are also in descending order.  
Hence, we get that $\hat{\vect{v}}=\bigvee_{v\in R}\vect{v}$. Furthermore, it 
is easy to check that $\hat{v}\in R$. Hence, the solution of the optimizations 
in Eqs.~\eqref{eq:Cl1fromCl2} and \eqref{eq:Cl1fromCl2simplified}, denoted 
$f(C_{l_2},\vect{d})$, is given by 
$\sqrt{2C_{l_2}}\sum_{k=1}^{n(n-1)/2}\sqrt{\hat{v}_k}$. Additionally, it is 
easy to check that, for fixed $\vect{d}$, $f(C_{l_2},\vect{d})$ is an 
increasing function on $C_{l_2}$. Combining these results with 
Eq.~\eqref{eq:estimationCl2}, we get the following result for estimating the 
$l_1$ norm of coherence,
\begin{equation}
  C_{l_1}(\rho)\ge f(S_L(\vect{d})-S_L(\vect{d}\vee\vect{p}),\vect{d}).
  \label{eq:estimationCl1}
\end{equation}

Third, we can also estimate the robustness of coherence from $C_{l_2}$ and 
$\vect{d}$. The robustness of coherence is defined as
\begin{equation}
  C_R(\rho)=\min\{r-1\mid\rho\le r\delta,~\delta\in\mathcal{I}\},
  \label{eq:robustnessCoherence}
\end{equation}
where $\mathcal{I}$ is the set of incoherent states. In addition, $C_R(\rho)$ 
also admits the dual form \cite{Napoli.etal2016}
\begin{equation}
  \begin{aligned}
    &\maxover[X]\quad        && \Tr(\rho X)-1\\
    &\subto                  && \Delta(X)=\I\\
    &                        && X\ge 0,
  \end{aligned}
  \label{eq:robustnessDual}
\end{equation}
where $\Delta(X)$ is the diagonal part of $X$. Without loss of generality we 
assume that $\rho_d$ is of full rank; otherwise, we only need to consider the 
coherence in the support of $\rho_d$. By taking 
$X=\rho_d^{-\frac{1}{2}}\rho\rho_d^{-\frac{1}{2}}$, which satisfies the 
constraints in Eq.~\eqref{eq:robustnessDual}, we get the inequality
\begin{equation}
  C_R(\rho)\ge\sum_{i\ne j}\frac{\abs{\rho_{ij}}^2}{\sqrt{d_id_j}},
  \label{eq:lowerBoundRobustness}
\end{equation}
which can be viewed as an improved estimation of that in 
Ref.~\cite{Napoli.etal2016}. Now, we consider the optimization problem
\begin{equation}
  \begin{aligned}
    &\minover[\rho_{ij}]\quad   && 2\sum_{i<j}\frac{\abs{\rho_{ij}}^2}
				{\sqrt{d_id_j}}\\
    &\subto                  && 2\sum_{i<j}\abs{\rho_{ij}}^2=C_{l_2}\\
    &                        && \abs{\rho_{ij}}^2\le d_id_j.
  \end{aligned}
  \label{eq:robustnessFromCl2}
\end{equation}
To get an analytic solution, we again denote 
$(2\abs{\rho_{ij}}^2/C_{l_2})_{i<j}$ and $(2d_id_j/C_{l_2})_{i<j}$  as 
$\vect{v}=(v_k)_{k=1}^{n(n-1)/2}$ and $\vect{u}=(u_k)_{k=1}^{n(n-1)/2}$, 
respectively, and assume that $u_k$ are in descending order. Then 
Eq.~\eqref{eq:robustnessFromCl2} turns to a linear program, and one can 
directly check that the optimal solution of Eq.~\eqref{eq:robustnessFromCl2}, 
which we denote $g(C_{l_2},\vect{d})$, is also attained when 
$\vect{v}=\hat{\vect{v}}$, where $\hat{\vect{v}}$ is given by 
Eq.~\eqref{eq:Cl1Join}. Again, for fixed $\vect{d}$, $g(C_{l_2},\vect{d})$ is 
an increasing function on $C_{l_2}$.  Combining these results with 
Eqs.~\eqref{eq:estimationCl2} and \eqref{eq:lowerBoundRobustness}, we get the 
following result for estimating the robustness of coherence,
\begin{equation}
  C_R(\rho)\ge g(S_L(\vect{d})-S_L(\vect{d}\vee\vect{p}),\vect{d}).
  \label{eq:estimationRobustness}
\end{equation}

Fourth, we consider two coherence measures based on the max entropy, $C_{\max}$ 
and $C_{\Delta,\max}$, which play crucial roles in one-shot coherence 
manipulation \cite{Zhao.etal2018,Regula.etal2018,Fang.etal2018,Zhao.etal2018b}.  
The coherence measures $C_{\max}$ and $C_{\Delta,\max}$ are defined as
\begin{equation}
  \begin{aligned}
    C_{\max}(\rho)&=\log_2\min\{r\mid\rho\le r\delta,~\delta\in\mathcal{I}\},\\
    C_{\max,\Delta}(\rho)&=\log_2\min\{r\mid\rho\le r\rho_d\}.
  \end{aligned}
  \label{eq:maxEntropyCoherence}
\end{equation}
From the definition, we can easily see that 
$C_{\max}(\rho)=\log_2(1+C_R(\rho))$.  Then we can get the estimation of 
$C_{\max}(\rho)$ directly from the estimation of $C_R(\rho)$ in 
Eq.~\eqref{eq:estimationRobustness}. In the following, we will focus on the 
estimation of $C_{\max,\Delta}(\rho)$.  Consider the optimization problem,
\begin{equation}
  C_{R,\Delta}(\rho)=\min\{r-1\mid\rho\le r\rho_d\},
  \label{eq:robustnessDeltaCoherence}
\end{equation}
then we have that
\begin{equation}
  C_{\max,\Delta}(\rho)=\log_2(1+C_{R,\Delta}(\rho)).
  \label{eq:maxEntropyDeltaCoherence}
\end{equation}
By applying the duality of semidefinite programming 
\cite{Boyd.Vandenberghe2004}, we can show that the optimization in 
Eq.~\eqref{eq:robustnessDeltaCoherence} is equivalent to the dual form
\begin{equation}
  \begin{aligned}
    &\maxover[X]\quad        && \Tr((\rho-\rho_d)X)\\
    &\subto                  && -X+(\Tr(X\rho_d))\I\le\I.
  \end{aligned}
  \label{eq:robustnessDeltaDual}
\end{equation}
Again, we assume that $\rho_d$ is of full rank.  By taking  
$X=\rho_d^{-1}\rho\rho_d^{-1}/n$, which satisfies the constraint in 
Eq.~\eqref{eq:robustnessDeltaDual}, we get the inequality
\begin{equation}
  C_{R,\Delta}(\rho)\ge\frac{1}{n}\sum_{i\ne 
  j}\frac{\abs{\rho_{ij}}^2}{d_id_j}.
  \label{eq:lowerBoundRobustnessDelta}
\end{equation}
Now, we consider the optimization problem
\begin{equation}
  \begin{aligned}
    &\minover[\rho_{ij}]\quad   && \frac{2}{n}\sum_{i<j}
				\frac{\abs{\rho_{ij}}^2}{d_id_j}\\
    &\subto                     && 2\sum_{i<j}\abs{\rho_{ij}}^2=C_{l_2}\\
    &                           && \abs{\rho_{ij}}^2\le d_id_j.
  \end{aligned}
  \label{eq:robustnessDeltaFromCl2}
\end{equation}
We again denote $(2\abs{\rho_{ij}}^2/C_{l_2})_{i<j}$ and 
$(2d_id_j/C_{l_2})_{i<j}$  as $\vect{v}=(v_k)_{k=1}^{n(n-1)/2}$ and 
$\vect{u}=(u_k)_{k=1}^{n(n-1)/2}$, respectively, and assume that $u_k$ are in 
descending order. Then Eq.~\eqref{eq:robustnessDeltaFromCl2} turns to a linear 
program, and the optimal solution of Eq.~\eqref{eq:robustnessDeltaFromCl2}, 
which we denote $h(C_{l_2},\vect{d})$, is also attained when 
$\vect{v}=\hat{\vect{v}}$, where $\hat{\vect{v}}$ is given by 
Eq.~\eqref{eq:Cl1Join}. Then we get an estimation of $C_{R,\Delta}$,
\begin{equation}
  C_{R,\Delta}(\rho)\ge h(S_L(\vect{d})-S_L(\vect{d}\vee\vect{p}),\vect{d}),
  \label{eq:estimationRobustnessDelta}
\end{equation}
and the estimation of $C_{\max,\Delta}$ follows directly from 
Eq.~\eqref{eq:maxEntropyDeltaCoherence}.

Morover, we can also use the majorization relation directly to estimate 
$C_{R,\Delta}$ and $C_{\max,\Delta}$.  Suppose that the vectors 
$\vect{\lambda}$, $\vect{c}=\vect{p}\vee\vect{d}$, and $\vect{d}$ are the 
spectrum, estimation of the spectrum, and diagonal entries of $\rho$, 
respectively, and the components of these probability distributions are in 
descending order.  Then we have that 
$\vect{\lambda}\succ\vect{c}\succ\vect{d}$, which means
\begin{equation}
  \sum_{i=1}^k\lambda_i\ge\sum_{i=1}^kc_i\ge\sum_{i=1}^kd_i,
  \label{eq:majorizationMeaning}
\end{equation}
for all $1\le k\le n$. Let $P_k$ be the orthogonal projectors onto the 
eigenspaces corresponding to the largest $k$ eigenvalues. Then $\rho\le 
r\rho_d$ implies that
\begin{equation}
  \sum_{i=1}^k\lambda_i=\Tr(P_k\rho)\le r\Tr(P_k\rho_d)\le r\sum_{i=1}^kd_i,
  \label{eq:projectionOrder}
\end{equation}
which gives us the bound
\begin{equation}
  C_{R,\Delta}(\rho)\ge\max_k\frac{\sum_{i=1}^k\lambda_i}{\sum_{i=1}^kd_i}-1
  \ge\max_k\frac{\sum_{i=1}^kc_i}{\sum_{i=1}^kd_i}-1.
  \label{eq:spectrumRobustness}
\end{equation}
Equation~\eqref{eq:spectrumRobustness} may provide a better estimation than 
Eq.~\eqref{eq:estimationRobustnessDelta} when $\Tr\rho^2$ is small, and again
the estimation of $C_{\max,\Delta}$ follows directly from 
Eq.~\eqref{eq:maxEntropyDeltaCoherence}.

Finally, we note that all these estimations can be naturally generalized to the 
multi-partite case, in which we only need to replace $\vect{p}$ with 
$\bigwedge_{\vect{p}\in X}\vect{p}$. All the bounds in 
Eqs.~\eqref{eq:estimationCl2}, \eqref{eq:estimationCl1}, 
\eqref{eq:estimationRobustness}, \eqref{eq:estimationRobustnessDelta}, and 
\eqref{eq:spectrumRobustness} are strictly positive when 
$\vect{p}\nprec\vect{d}$. The bounds in Eqs.~\eqref{eq:estimationCl2}, 
\eqref{eq:estimationCl1}, \eqref{eq:estimationRobustness}, and 
\eqref{eq:estimationRobustnessDelta} can be attained when the states are pure.  
The bound in Eq.~\eqref{eq:spectrumRobustness} can be attained when $\rho_d$ 
are maximally mixed states.

\section{Construction of the linear program}
\label{sec:example}

To make the process of constructing the linear program more concrete, we 
explicitly show how to construct the constraints for $N=3$ in this Appendix.  
According to the discussion in the text, we can get mean values of the 
observables
\begin{equation}
  \begin{aligned}
    M_0&=\I\otimes\I\otimes\I,\\
    M_1&=Z\otimes\I\otimes Z,\\
    M_2&=Z\otimes Z\otimes\I,\\
    M_3&=\I\otimes Z\otimes Z,\\
    M_4&=X\otimes X\otimes X.
  \end{aligned}
  \label{eq:halfStablizer}
\end{equation}
from the measurement settings $\mathcal{X}$ and $\mathcal{Z}$, where 
$\mean{M_0}=1$ just denotes the normalization of the probability distribution.
Furthermore, the observables in Eq.~\eqref{eq:halfStablizer} are diagonal in 
the GHZ basis,
\begin{equation}
  \begin{aligned}
    \ket{\psi_0}=\frac{1}{\sqrt{2}}(\ket{000}+\ket{111}),~~
    \ket{\psi_4}=\frac{1}{\sqrt{2}}(\ket{000}-\ket{111}),\\
    \ket{\psi_1}=\frac{1}{\sqrt{2}}(\ket{001}+\ket{110}),~~
    \ket{\psi_5}=\frac{1}{\sqrt{2}}(\ket{001}-\ket{110}),\\
    \ket{\psi_2}=\frac{1}{\sqrt{2}}(\ket{010}+\ket{101}),~~
    \ket{\psi_6}=\frac{1}{\sqrt{2}}(\ket{010}-\ket{101}),\\
    \ket{\psi_3}=\frac{1}{\sqrt{2}}(\ket{011}+\ket{100}),~~
    \ket{\psi_7}=\frac{1}{\sqrt{2}}(\ket{011}-\ket{100}).
  \end{aligned}
  \label{eq:GHZbasis}
\end{equation}
Explicitly, we have
\begin{equation}
  M_i=\sum_{j=0}^7B_{ij}\ket{\psi_j}\bra{\psi_j},
  \label{eq:eigenDecomposition}
\end{equation}
for $i=0,1,\dots,4$, where
\begin{equation}
  B=
  \begin{bmatrix}
    H\otimes H & H\otimes H\\
    \vect{1}^T & -\vect{1}^T
  \end{bmatrix},
  \label{eq:explicitB}
\end{equation}
with
\begin{equation}
  \begin{aligned}
    H&=
    \begin{bmatrix}
      1 & 1\\
      1 & -1
    \end{bmatrix},\\
    \vect{1}^T&=[1,1,1,1].
  \end{aligned}
  \label{eq:Hadamard}
\end{equation}
Let $p_j$ denote the probabilities in the GHZ basis, i.e.,
\begin{equation}
  p_j=\bra{\psi_j}\rho\ket{\psi_j},
  \label{eq:GHZprobabilities}
\end{equation}
for $i=0,1,\dots,7$. Then we get that
\begin{equation}
  \beta_i\equiv\mean{M_i}=\Tr(M_i\rho)=\sum_{j=0}^7B_{ij}p_j,
  \label{eq:explicitConstrints}
\end{equation}
for $i=0,1,\dots,4$, which are just the linear inequality constraints 
$B\vect{p}=\vect{\beta}$. The linear inequality constraints are just 
$\vect{p}\ge 0$, i.e., $A=\I$ and $\vect{\alpha}=\vect{0}$. Note that we do not 
assume that state $\rho$ is diagonal in the GHZ basis in the derivation.  
Hence, the estimation works for any state $\rho$, not just the GHZ-diagonal 
ones.

\section{Symmetrizing the optimization}
\label{sec:symmetrization}

The main factor that affects the performance of the linear program for 
$\bigwedge_{\vect{p}\in X}\vect{p}$ is the number of equality and inequality 
constraints in $X$. In our example for the GHZ basis, there are $2^{N-1}+1$ 
equality constraints and $2^N$ inequality constraints in $X$ for an $N$-qubit 
quantum system. This is still tractable for up to $20$ qubits in standard 
computers.  It may be beyond the capability of the current hardware when $N$ 
goes to $30$ or even larger. In this Appendix, we propose a method to overcome 
this by using the symmetrization technique \cite{Toth.etal2010}. We use the 
permutation symmetry to reduce both the number of constraints and the number of 
optimization variables. The final symmetrized linear program is only $O(N)$ for 
an $N$-qubit quantum system. Moreover, the solution of the symmetrized linear 
program is exactly the same as the original one if the quantum state is 
symmetric under the permutations of the qubits.

Before describing the method, we first define some notations. We use $S_N$ to 
denote the symmetric group of the $N$ qubits and $\mathcal{S}$ to denote the 
symmetrization operator that makes an expression invariant under $S_N$. For 
example, for $N=3$,
\begin{equation}
  \begin{aligned}
    &\mathcal{S}(Z\otimes Z\otimes\I)=Z\otimes Z\otimes\I+ Z\otimes\I\otimes 
    Z+\I\otimes Z\otimes Z,\\
    &\mathcal{S}(X\otimes X\otimes X)=X\otimes X\otimes X,\\
    &\mathcal{S}(p_{001}^\pm)=\mathcal{S}(p_{010}^\pm)=\mathcal{S}(p_{011}^\pm)
    =p_{001}^\pm+p_{010}^\pm+p_{011}^\pm,\\
    &\mathcal{S}(p_{000}^\pm)=p_{000}^\pm.
  \end{aligned}
  \label{eq:symmEx}
\end{equation}

Now, we define the symmetrized optimization problem for majorization join; we 
replace the linear equality constraints induced by $\mean{\mathcal{Z}_E}$ by 
$\mean{\mathcal{S}(\mathcal{Z}_E)}$ and the inequality constraints $p_l^\pm\ge 
0$ with $\mathcal{S}(p_l^\pm)\ge 0$. It is easy to check that the symmetrized 
optimization problem,
\begin{equation}
  \begin{aligned}
    &\minover[\vect{p}]\quad && \sum_{i=1}^kp_i^{\downarrow}\\
    &\subto                  && \tilde{A}\vect{p}\ge\tilde{\vect{\alpha}}\\
    &                        && \tilde{B}\vect{p}=\tilde{\vect{\beta}},
  \end{aligned}
  \label{eq:symmMeetPrimal}
\end{equation}
has $\lceil (N+1)/2\rceil+1$ equality constraints and $2\lfloor N/2\rfloor+2$ 
inequality constraints.  Let 
$\tilde{X}=\{\vect{p}\mid\tilde{A}\vect{p}\ge\tilde{\vect{\alpha}},~
\tilde{B}\vect{p}=\tilde{\vect{\beta}}\}$. Then we have $X\subset\tilde{X}$.  
This implies that $\bigwedge_{\vect{p}\in\tilde{X}}\vect{p}\prec 
\bigwedge_{\vect{p}\in X}\vect{p}\prec\vect{\lambda}$. Hence, the optimization 
in Eq.~\eqref{eq:symmMeetPrimal} provides a relaxation of the original 
optimization problem in Eq.~\eqref{eq:meetPrimalA}.

Then, we show that the solution of the symmetrized optimization in 
Eq.~\eqref{eq:symmMeetPrimal} is exactly the same as the original one in 
Eq.~\eqref{eq:meetPrimalA}, if the quantum state is symmetric under $S_N$. To 
this end, we first show that in the symmetrized optimization in 
Eq.~\eqref{eq:symmMeetPrimal}, we only need to consider the set of all 
symmetric probability distributions $S$, i.e., the probability distributions 
that are invariant under $S_N$. That is we need to prove that \begin{equation}
  \textstyle{\bigwedge}_{\vect{p}\in\tilde{X}}\vect{p}=
  \textstyle{\bigwedge}_{\vect{p}\in\tilde{X}\cap S}\vect{p}.
  \label{eq:symmEquivalent}
\end{equation}
On one hand, $\bigwedge_{\vect{p}\in\tilde{X}}\vect{p}\prec 
\bigwedge_{\vect{p}\in\tilde{X}\cap S}\vect{p}$ is obvious from the definition 
of the majorization meet. On the other hand, for any $\vect{p}\in\tilde{X}$, we 
have $\pi(\vect{p})\in\tilde{X}$ for all $\pi\in S_N$, because $\tilde{X}$ is 
symmetric under $S_N$. Then the average of $\vect{p}$ under $S_N$ is symmetric 
and in $\tilde{X}$, i.e., $\bar{\vect{p}}=\frac{1}{n!}\sum_{\pi\in 
S_N}\pi(\vect{p})\in\tilde{X}\cap S$. Furthermore, we have 
$\bar{\vect{p}}\prec\vect{p}$ \cite{Bengtsson.Zyczkowski2017}. This implies 
that $\bigwedge_{\vect{p}\in\tilde{X}\cap S}\vect{p}= 
\bigwedge_{\vect{p}\in\tilde{X}}\bar{\vect{p}}\prec
\bigwedge_{\vect{p}\in\tilde{X}}\vect{p}$. Thus, we prove 
Eq.~\eqref{eq:symmEquivalent}. With the same technique, we can also prove that 
if $\mean{\mathcal{Z}_E}=\mean{\pi(\mathcal{Z}_E)}$ for all $\pi\in S_N$, then
\begin{equation}
  \textstyle{\bigwedge}_{\vect{p}\in X}\vect{p}=
  \textstyle{\bigwedge}_{\vect{p}\in X\cap S}\vect{p},
  \label{eq:nonsymmEquivalent}
\end{equation}
for the optimization in Eq.~\eqref{eq:meetPrimalA}. Furthermore, we have 
$\tilde{X}\cap S=X\cap S$ in this case. This implies that the solution of the 
symmetrized optimization in Eq.~\eqref{eq:symmMeetPrimal} is exactly the same 
as the original one in Eq.~\eqref{eq:meetPrimalA}, if the quantum state is 
symmetric under $S_N$.

Finally, we show how to further simplify the corresponding linear program by 
taking advantage of the permutation symmetry. As we have shown above, we only 
need to consider the symmetrized probabilities $\mathcal{S}(p_l^\pm)$, which 
only have $2\lfloor N/2\rfloor+2$ different ones. We denote these $2\lfloor 
N/2\rfloor+2$ probabilities $\vect{s}=(s_i^\pm)_{i=0}^{\lfloor N/2\rfloor}$, 
where $s_i^\pm$ is the sum of $c_i^\pm$ elements in $\vect{p}=(p_l^{\pm})$ with
\begin{equation}
  c_i^\pm=
  \begin{cases}
    \binom{N-1}{i}+\binom{N-1}{N-i}=\binom{N}{i}
    &\text{for }i\ne\frac{N}{2},\\[1em]
    \binom{N-1}{i}=\frac{1}{2}\binom{N}{i}
    &\text{for }i=\frac{N}{2}.
  \end{cases}
  \label{eq:coefficients}
\end{equation}
Let $\tilde{Y}_k=\{\vect{y}\mid\vect{0}\le\vect{y}\le\vect{1},~
\vect{c}^T\vect{y}=k\}$; then we have 
$\sum_{i=1}^kp_i^{\downarrow}=\max_{\vect{y}\in 
\tilde{Y}_k}\vect{y}^T\vect{c}$, where $\vect{c}=(c_i^\pm)_{i=0}^{\lfloor 
N/2\rfloor}$. Let $\vect{p}=\tilde{C}\vect{s}$
for $\vect{p}\in\tilde{X}\cup S$; then the constraint in $\tilde{X}$ turns to
$\tilde{X}_s=\{\vect{s}\mid\tilde{A}\tilde{C}\vect{s}\ge\tilde{\alpha},~
\tilde{B}\tilde{C}\vect{s}=\tilde{\beta}\}$. Thus we get $\min_{\vect{p}\in
\tilde{X}}\sum_{i=1}^kp_i^{\downarrow}=\min_{\vect{p}\in
\tilde{X}\cup S}\sum_{i=1}^kp_i^{\downarrow}=\min_{\vect{s}\in\tilde{X}_s}
\max_{\vect{y}\in\tilde{Y}_k}\vect{y}^T\vect{s}$. Following the same argument 
as in Appendix \ref{sec:adaptive}, we can prove that the symmetrized 
optimization in Eq.~\eqref{eq:symmMeetPrimal} is equivalent to the symmetrized 
linear program,
\begin{equation}
  \begin{aligned}
    &\maxover[\vect{\mu},\vect{\nu}]\quad && \vect{\tilde{\alpha}}^T\vect{\mu}
			 +\vect{\tilde{\beta}}^T\vect{\nu}\\
    &\subto &&
    \vect{0}\le\tilde{C}^T\tilde{A}^T\vect{\mu}
			 +\tilde{C}^T\tilde{B}^T\vect{\nu}\le\vect{1}\\
    &       &&
    \vect{c}^T\tilde{C}^T\tilde{A}^T\vect{\mu}
			 +\vect{c}^T\tilde{C}^T\tilde{B}^T\vect{\nu}=k\\
    &       && \vect{\mu}\ge\vect{0},
  \end{aligned}
  \label{eq:symmMeetDual}
\end{equation}
which is only $O(N)$ for an $N$-qubit quantum system.

\section{Characterizing the freezing of coherence}
\label{sec:freezing}

In this Appendix, we discuss some details about the model for characterizing 
the freezing of coherence. The channel we consider is the local bit-flip 
channel $\Lambda^{\otimes N}$, where $\Lambda$ satisfy the Lindblad equation,
\begin{equation}
  \frac{\d}{\d t}\Lambda(\rho)=\frac{\gamma}{2}(\sigma_x\rho\sigma_x-\rho),
  \label{eq:Lindblad}
\end{equation}
and $\gamma$ is a parameter that represents the strength of the noise.
The solution of Eq.~\eqref{eq:Lindblad} is given by
\begin{equation}
  \Lambda(\rho)=\frac{1}{2}(1+e^{-\gamma t})\rho+\frac{1}{2}(1-e^{-\gamma 
  t})\sigma_x\rho\sigma_x.
  \label{eq:bitflip}
\end{equation}
The initial state we consider is the $N$-qubit GHZ state, where the preparation 
of the initial state is inevitably affected by noise in experiments. We 
consider the two most common types of noise in experiments, dephasing noise 
$\Delta_\varepsilon$ and depolarizing noise $\mathcal{D}_\varepsilon$,
\begin{equation}
  \begin{aligned}
    &\text{Dephasing:}\quad 
    &&\Delta_\varepsilon(\rho)=(1-\varepsilon)\rho+\varepsilon\rho_d,\\
    &\text{Depolarizing:}\quad
    &&\mathcal{D}_\varepsilon(\rho)=(1-\varepsilon)\rho+\varepsilon\frac{\I}{2}.
  \end{aligned}
  \label{eq:noise}
\end{equation}
For the $N$-qubit GHZ state, the initial state affected by dephasing noise is 
given by
\begin{equation}
  \rho_0^\Delta=\frac{1}{2}(1+(1-\varepsilon)^N)
  \ket{\varphi_{0\dots 0}^+}\bra{\varphi_{0\dots 0}^+}
  +\frac{1}{2}(1-(1-\varepsilon)^N)
  \ket{\varphi_{0\dots 0}^-}\bra{\varphi_{0\dots 0}^-},
  \label{eq:dephased}
\end{equation}
and the initial state affected by depolarizing noise is given by
\begin{equation}
  \begin{aligned}
    \rho_0^\mathcal{D}&=\frac{1}{2}((1-\frac{\varepsilon}{2})^N
      +(\frac{\varepsilon}{2})^N+(1-\varepsilon)^N)
      \ket{\varphi_{0\dots 0}^+}\bra{\varphi_{0\dots 0}^+}\\
      &+\frac{1}{2}((1-\frac{\varepsilon}{2})^N
      +(\frac{\varepsilon}{2})^N-(1-\varepsilon)^N)
      \ket{\varphi_{0\dots 0}^-}\bra{\varphi_{0\dots 0}^-}\\
      &+\sum_{w(l)\ne 0,N}\frac{1}{2}((1-\frac{\varepsilon}{2})^{w(l)}
	(\frac{\varepsilon}{2})^{N-w(l)}+(\frac{\varepsilon}{2})^{w(l)}
      (1-\frac{\varepsilon}{2})^{N-w(l)})\ket{l}\bra{l},
  \end{aligned}
  \label{eq:3qubit}
\end{equation}
where $w(l)$ is the Hamming weight of the binary string $l=l_1l_2\dots l_N$, 
and
\begin{equation}
  \ket{\varphi_{0\dots 0}^\pm}=\frac{1}{\sqrt{2}}(\ket{0\dots 
  0}\pm\ket{1\dots 1}).
  \label{eq:GHZ0}
\end{equation}
The fidelities of the initial states affected by dephasing noise and 
depolarizing noise are given by
\begin{equation}
  \begin{aligned}
    F^{\Delta}&=\frac{1}{2}(1+(1-\varepsilon)^N),\\
    F^{\mathcal{D}}&=\frac{1}{2}((1-\frac{\varepsilon}{2})^N
    +(\frac{\varepsilon}{2})^N+(1-\varepsilon)^N),
  \end{aligned}
  \label{eq:fidelities}
\end{equation}
respectively.


\begin{acknowledgments}
  We would like to thank Mariami Gachechiladze, Barbara Kraus, Katharina 
  Schwaiger, and Nikolai Wyderka for discussions. We also thank the anonymous 
  referees for many helpful suggestions. This work was supported by the DFG and 
  the ERC (Consolidator Grant 683107/TempoQ). X.D.Y. acknowledges funding from 
  a CSC-DAAD scholarship.
\end{acknowledgments}

\bibliography{QuantumInf}

\end{document}